\documentclass[reprint, amssymb, amsmath, amsfonts, a4paper, aps, floatfix, superscriptaddress, showpacs]{revtex4-1}
\usepackage{tabularx, xspace, rotating, units, enumerate, ucs}

\usepackage[modulo, switch]{lineno}

\newcommand{\Epi}{\affiliation{Department of Epileptology, University of Bonn, Sigmund-Freud-Stra{\ss}e~25, 53105~Bonn, Germany}}
\newcommand{\HISKP}{\affiliation{Helmholtz Institute for Radiation and Nuclear Physics, University of Bonn, Nussallee~14--16, 53115~Bonn, Germany}}
\newcommand{\IZKS}{\affiliation {Interdisciplinary Center for Complex Systems, University of Bonn, Br\"uhler Stra\ss{}e~7, 53175~Bonn, Germany}}
\newcommand{\ICBM}{\affiliation {Theoretical Physics/Complex Systems, ICBM, Carl von Ossietzky University of Oldenburg, \\Carl-von-Ossietzky-Stra\ss{}e~9--11, Box~2503, 26111~Oldenburg, Germany}}
\newcommand{\RNS}{\affiliation{Research Center Neurosensory Science, Carl von Ossietzky University of Oldenburg,\\ Carl-von-Ossietzky-Stra\ss{}e~9--11, 26111~Oldenburg, Germany}}
\newcommand{\Maryland}{\affiliation{Institute for Physical Science and Technology, University of Maryland, College Park, MD 20742-2431, U.S.A.}}

\newcommand{\kl}[1]{\left( #1 \right)}
\newcommand{\klg}[1]{\left\{ #1 \right\}}

\newcommand{\defi}{\mathrel{\mathop:}=}

\newcommand*\diff{\mathop{}\!\mathrm{d}}

\newcommand{\lkl}{\lessapprox k \lessapprox }

\begin{document}

\title{Route to extreme events in excitable systems}

\author{Rajat Karnatak}
\ICBM

\author{Gerrit Ansmann}
\Epi \HISKP \IZKS

\author{Ulrike Feudel}
\ICBM \RNS \Maryland

\author{Klaus Lehnertz}
\Epi \HISKP \IZKS

\begin{abstract}
Systems of FitzHugh--Nagumo units with different coupling topologies are capable of self-generating and \nobreakdash-terminating strong deviations from their regular dynamics that can be regarded as extreme events due to their rareness and recurrent occurrence.
Here we demonstrate the crucial role of an interior crisis in the emergence of extreme events.
In parameter space we identify this interior crisis as the organizing center of the dynamics by employing concepts of mixed-mode oscillations and of leaking chaotic systems.
We find that extreme events occur in certain regions in parameter space, and we show the robustness of this phenomenon with respect to the system size.
\end{abstract}

\pacs{05.45.Xt 89.75.-k 89.75.Fb
}
\maketitle

\section{Introduction}
Extreme events have been a topic of increasing interest during the last decade~\cite{Albeverio2006, Ghil2011, Smith2013} and occur in various contexts like in geophysics~\cite{Sornette1996, Douglas1999, Stark2001, Oppenheimer2003, Mcdaniels2008}, meteorology~\cite{Peterson2013}, economics~\cite{Bordo1986, Jacobs1999}, power and communication grids~\cite{Labovitz1999, Kinney2005, Sun2005, Kiciman2005}.
Many of these works have been devoted to the specific statistical properties of such events.
Here we take a different point of view and emphasize the perspective of dynamical systems.
We define an extreme event as a rare and recurrent event on which an appropriate variable exhibits an unusual behavior, e.g., possesses an extremely large or small value~\cite{Ansmann2013}.
This definition does not require the events to conform to a certain statistics.
Whether an extreme event can be observed depends crucially on the observable chosen:
Simple examples of appropriate observables in physical systems would be the wave height for oceanic waves~\cite{Kharif2009} or the intensity of the optical output for optical rogue waves~\cite{Solli2007, Bonatto2011}.
Less simple observables, because of not being obvious, would be the abundance of a toxic algal species in a harmful algal bloom~\cite{Anderson2001} since it may not be the most abundant species in the plankton community, but have the largest impact on the ecosystem due to their toxin production.
Another example would be the level of synchrony of populations of neurons in the brain~\cite{Lehnertz2006}, which could be considered as an appropriate indication for an epileptic seizure, which is an extreme event to the affected person.
The two latter examples illustrate that the observable used to demonstrate an extreme event is specific to the application.

Different mechanisms behind the appearance of extreme events have been discussed in the literature.
In the complex Ginzburg--Landau equation, instabilities lead to the formation of a weakly interacting and incoherent background of low-amplitude waves, which under certain conditions can collapse locally yielding a large-amplitude event~\cite{Kim2003}.
In arrays of locally coupled lasers with randomly distributed natural frequencies, wandering localized excitations resulting from a progressive spatial synchronization of the lasers with an increasing coupling strength have been observed~\cite{Rogister2007}.
In a multistable laser system, noise-induced attractor hopping has been proposed as the mechanism behind the appearance of optical rogue waves~\cite{Pisarchik2011}.
Recently, chimera states in small-world networks of pulse-coupled oscillators have been shown to correspond to events of extreme synchrony under certain conditions~\cite{Rothkegel2014}.

In our previous paper~\cite{Ansmann2013}, we have demonstrated that systems of FitzHugh--Nagumo units \cite{VanDerPol1928, Bonhoeffer1948, FitzHugh1961, Nagumo1962} with different coupling topologies are capable of self-generating and \nobreakdash-terminating extreme events in the above sense, and we described their dynamical properties and underlying basic mechanisms.
This study was performed only with a particular set of control parameters.
The goal of the present paper is to discuss the mechanisms behind these events in more detail by analyzing the parameter space, identifying the general properties of the transition to the emergence of extreme events and investigating the robustness of this phenomenon.
We will illustrate the role of an interior crisis in which the period-doubling cascade from one side of the parameter space collides with the period-adding cascade from the other~\cite{Fan1995}.
We will employ the concept of mixed-mode oscillations~\cite{Desroches2012} to analyze the behavior in parameter space and the theory of leaking chaotic systems~\cite{Altmann2013} to discuss the frequency of extreme events.

The paper is arranged as follows: In Sec.~\ref{sec:systems}, we introduce the model systems and briefly recapitulate our results from Ref.~\cite{Ansmann2013}.
In Sec.~\ref{sec:ob2coupled} we analyze and discuss the intricate structure of the parameter dependencies of the two-unit system from several perspectives, namely those of bifurcations in which chaotic dynamics with immersed extreme events and mixed-mode oscillations alternate.
We investigate to which extent these observations carry over to the case of 101 globally coupled units and discuss the robustness of the appearance of extreme events in Sec.~\ref{sec:ob.manycoupled}.
In Sec.~\ref{sec:dependence_systemsize}, we take a general look at the change of the dynamical regimes with the system size and draw the conclusions from our analysis in Sec.~\ref{sec:conclusions}.

\section{Model systems}\label{sec:systems}

We consider model systems as in Ref.~\cite{Ansmann2013}, namely $n$~FitzHugh--Nagumo units, which are coupled completely and diffusively and which are described by the following differential equations ($i \in \klg{1, \ldots, n}$):
\begin{align}
	\dot{x}_i & = x_i (a-x_i) (x_i-1) - y_i + K \sum\limits_{j=1}^{n} (x_j - x_i) \notag\\
	\dot{y}_i & = b_i x_i - c y_i \label{eq:systems}
\end{align}
Here, $a$, $b_i$, and~$c$ are internal parameters of the unit and $K$ denotes the coupling strength.
In the following, we consider $k\defi K (n-1)$ for a better comparability of differently sized systems.
For a given system, $a$~and~$c$ are identical for all units, while $b$ is mismatched.

In particular, we regard the following two systems:

(A)	A system of $n=2$ units with $a=-0.025794$, $c=0.02$, $b_1=0.0135$, and $b_2=0.0065$.

(B)	A system of $n=101$ units with $a=-0.02651$, $c=0.02$ and $b_i = 0.006 + \tfrac{i-1}{n-1}\cdot 0.008$ ($\Rightarrow 0.006 \leq b_i \leq 0.014 \,\forall\,i$).

Note that the parameters $a$, $b$, and $c$ were chosen comparably or identically for both systems.

\begin{figure}
\includegraphics{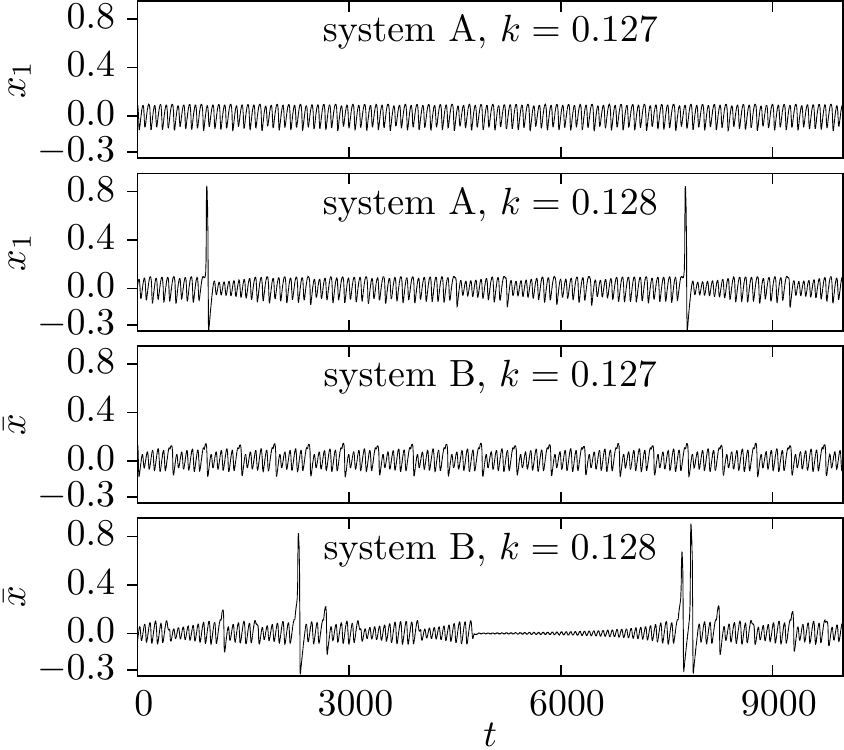}
	\caption{
		Exemplary temporal evolutions of both model systems for different coupling strengths.
		For consistency with the continuation calculations (cf. Fig.~\ref{fig:bifurcation2}), we use $x_1$ instead of $\bar{x}$ as the main observable for system~A.
	}
	\label{fig:timeseries}
\end{figure}

Both systems were realized with 4\textsuperscript{th}-order Runge--Kutta methods, either with a fixed step size of $0.01$ or an adaptive step size with a maximum estimated relative error of $10^{-5}$ (Runge--Kutta--Fehlberg, realized with Conedy~\cite{Rothkegel2012}).

For a certain set of parameter values, the average of the first dynamical variable of these systems $\bar{x}\kl{t} \defi \tfrac{1}{n} \sum_{i=1}^n x_i\kl{t}$ exhibits rare events of high amplitude, which are extreme events in our understanding (see the second and fourth row of Fig.~\ref{fig:timeseries}).
During such an event, all units of the system become excited simultaneously.
We observed for system~A that extreme events occur due to a \textit{channel-like structure} in state space, which exists due to the alignment of the manifolds of the saddle focus at the origin.
For system~B, we found that it frequently exhibits \textit{proto-events,} during which a fraction of the units become excited and which turn into extreme events, if and only if this fraction is sufficiently large.
We here will take a closer look at changes of the system dynamics with these parameters.
Since these changes do not strongly depend on which parameter is varied, we focus on the coupling strength~$k$.
Changes of internal parameters are discussed in Appendix~\ref{sec:internal}.
Our main focus lies in the study of the emergence of extreme events in parameter space and we particularly address the robustness of this phenomenon.

For each of the following observations and analyses, at least 5000 initial time units were discarded.
The initial conditions were chosen randomly and had no influence on our observations, unless noted otherwise.
To facilitate automatized detection, we define an extreme event as a time interval with $\bar{x}\kl{t}>0.4$ and consider the time difference between two of their subsequent beginnings as inter-event intervals.

\section{System~A}
\label{sec:ob2coupled}

In this section, we discuss the dynamical changes of two coupled FitzHugh--Nagumo units depending on the coupling strength~$k$ (system~A).
A small note regarding the representation:
An appropriately placed Poincar\'e section on a period\nobreakdash-$T$ limit cycle with one local maximum yields a period-one stable equilibrium of the corresponding map.
From this point of view, a period\nobreakdash-$T$ limit cycle with $m$~local maxima would be referred to as a \textit{period-$m$ attractor} or \textit{limit cycle} in the following.

\subsection{Bifurcation analysis}
\label{sec:bifurcation}

For $k=0$, the units exhibit high-amplitude oscillations.
With an increase of the coupling strength, the system shows a variety of complex dynamical behaviors, which include chaos.
Regimes of multistability are also observed; the system exhibits coexisting periodic--chaotic, and periodic--periodic, fixed-point--periodic--chaotic and fixed-point--periodic--periodic regimes in certain coupling intervals.
For completeness, a detailed discussion of the observed transitions for $0 < k \lessapprox 0.01218$ is presented in Appendix~\ref{sec:detailsoftransitions}.

For $k > 0.01218$, the system is in a regime where the equilibrium at the origin is the only dynamical attractor.
In the literature, this suppression of oscillations in nonlinear systems has been called \textit{amplitude} or \textit{oscillator death} \cite{BarEli1985, Mirollo1990b, Ermentrout1990}.
Given that our system features instantaneous coupling in identical variables, the suppression of oscillations in this system is a result of parameter mismatch between the coupled units~\cite{Aronson1990}.

\begin{figure}
\includegraphics{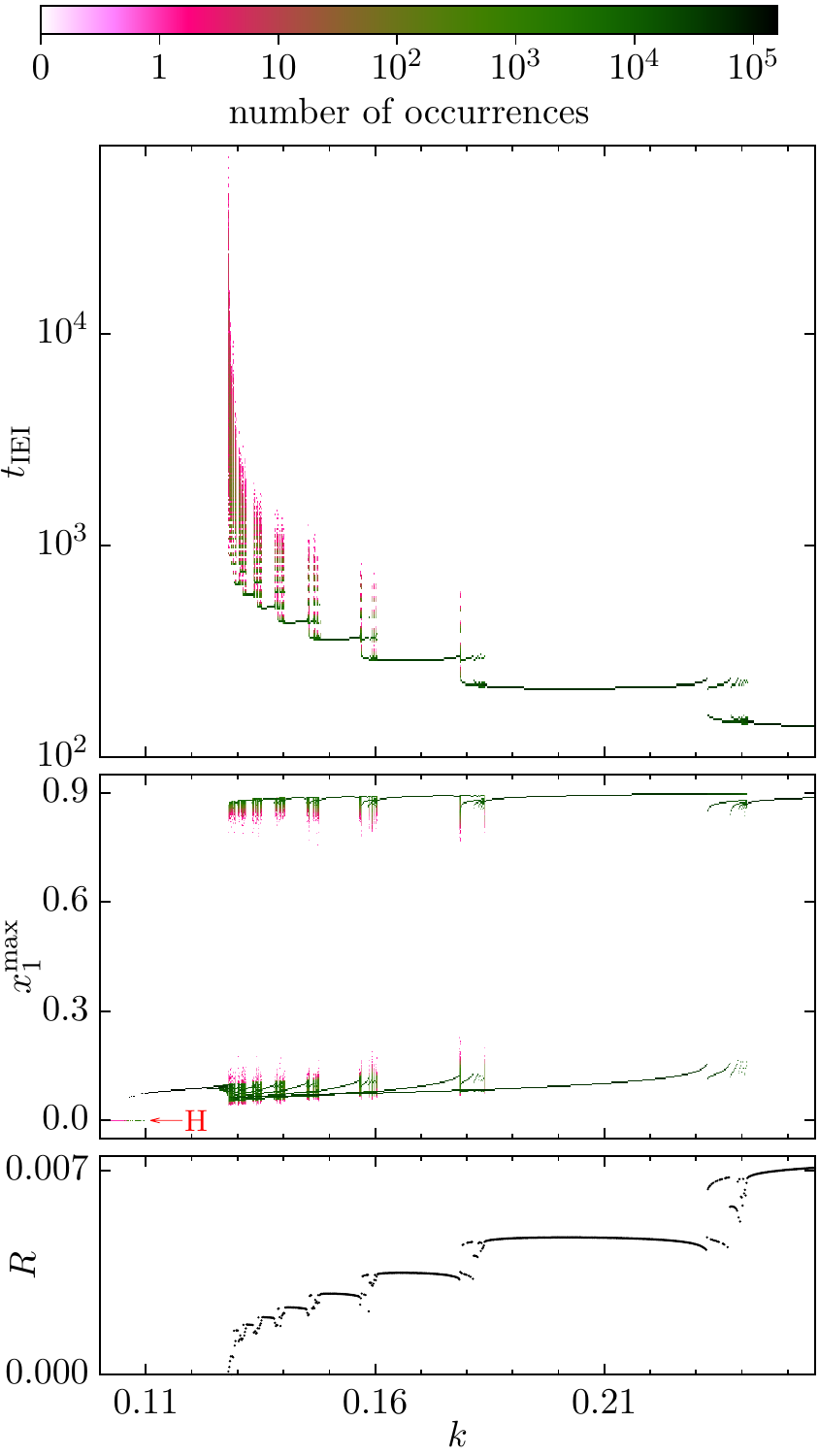}
	\caption{(Color online) (Top row) Inter-event intervals $t_\text{IEI}$ for system~A observed in 100000 time units in dependence of the coupling strength~$k$.
	For each $k$, the vertical slice of the diagram corresponds to a histogram with logarithmic bins, where the number of occurrences for each bin is color-coded.
	(Middle row) Same as top, but for local maxima $x_1^\text{max}$ of $x_1$, with a linear ordinate, and for 20 observations of 5000 time units each (different initial conditions).
	We consider a value $x_1(t)$ a local maximum, if $x_1(t-h)<x_1(t)>x_1(t+h)$, where $h$ is the sampling time.
	H~marks the Hopf bifurcation.
	(Bottom row): Event rate~$R$.
	}
	\label{fig:bifurcation1}
\end{figure}

\begin{figure}
\includegraphics{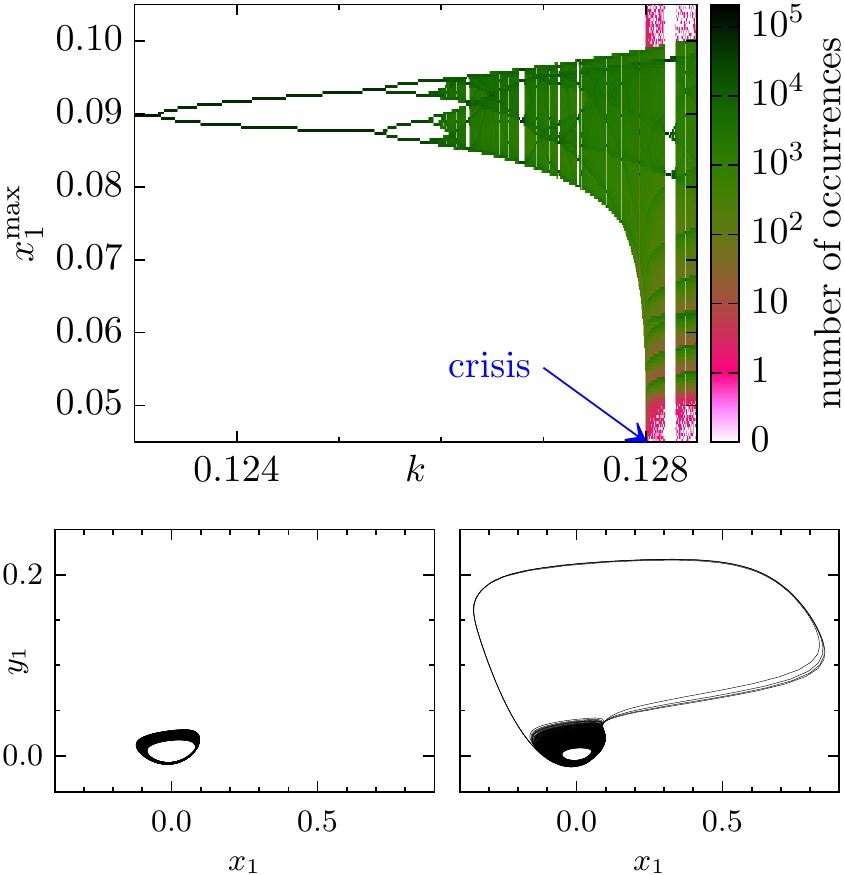}
	\caption{(Top) Detail of the second row of Fig.~\ref{fig:bifurcation1}.
	(Bottom) Attractor projection on the $(x_1,y_1)$ plane before ($k=0.127$, left) and after the interior crisis ($k=0.128$, right).}
	\label{fig:ba-crisis}
\end{figure}

The origin remains stable up to $k \approx 0.11068$, where a subcritical Hopf bifurcation gives rise to a period-one limit cycle (see the bifurcation diagrams in the top two parts of Fig.~\ref{fig:bifurcation1}).
This periodic behavior is stable for $0.10598 \lkl 0.12323$ and, starting from $k \approx 0.12321$, follows a period-doubling route to chaos (see Fig.~\ref{fig:ba-crisis}).
The resulting chaotic dynamics is interrupted by small periodic windows (e.g., for  $0.1267 \lkl 0.1268$) and undergoes an interior crisis slightly below $k = 0.128$.
Before the crisis point, the system evolves chaotically and is confined to a comparatively small part of the state space (cf. Fig.~\ref{fig:timeseries}, top and Fig.~\ref{fig:ba-crisis}, bottom left).
Beyond this point, the system still exhibits bounded chaotic behavior for the majority of its evolution, but aperiodically and rarely escapes from its confined region in state space and exhibits a long excursion (cf. Fig.~\ref{fig:timeseries}, second row and Fig.~\ref{fig:ba-crisis}, bottom right)---the extreme event.
After exhibiting this excursion, the system returns to its bounded chaotic oscillatory state until the next event.

The behavior of the system suggests that parts of the state space which were inaccessible to the trajectories before the crisis point have become accessible beyond it.
This is due to the opening of the channel-like structure in state space, through which the trajectories can escape.
The appearance of this structure can be interpreted as follows:
Since the Jacobian of the system depends on the coupling strength~$k$, so are the eigenvalues and the corresponding eigenvectors of the Jacobian.
Subsequently, the stable and unstable manifolds and their alignment are functions of~$k$ as well.
Before the crisis point, these manifolds are aligned such that they keep the trajectories bounded.
With changes of the coupling strength, this alignment also undergoes changes and with the interior crisis, the alignment leads to the opening of a gap, the channel-like structure.
Trajectories which venture into the part of the state space where this structure is located can escape through this gap for a long excursion.

If the coupling strength is increased beyond the crisis value, the system continues exhibiting chaotic behavior, but with extreme events.
The coupling regime for which chaotic behavior can be observed is interrupted by periodic windows.
The latter emerge from saddle-node bifurcations of limit cycles, remain stable for a certain range of coupling strengths, and lose stability via period doubling, giving rise to chaos again.
This pattern repeats with increasing~$k$, but the periodic windows increase in size, while the chaotic windows decrease until they completely vanish at $k \approx 0.241041$.
Importantly, in the chaotic windows, the event rate~$R$ (i.e., the number of events divided by the observation time) roughly increases with increasing~$k$ (see Fig.~\ref{fig:bifurcation1}, bottom).
We will discuss this in detail in Sec.~\ref{sec:channelproperties}.

We expect that the mechanism behind the appearance of the interior crisis is the collision of the different bifurcation processes of period doubling and reverse period adding.
In the literature, it has been proposed that a collision of two opposite bifurcation processes with different topological entropies \cite{Fan1995} can lead to an interior crisis, which appears to be the case here.

This analysis in parameter space shows that extreme events are not a phenomenon which is restricted to a particular point in parameter space, but which emerges in the chaotic regimes beyond the interior crisis, though the frequency of these events varies in and between the chaotic regions (see Fig.~\ref{fig:bifurcation1}, bottom).

\begin{figure*}
\includegraphics[width=0.7\linewidth]{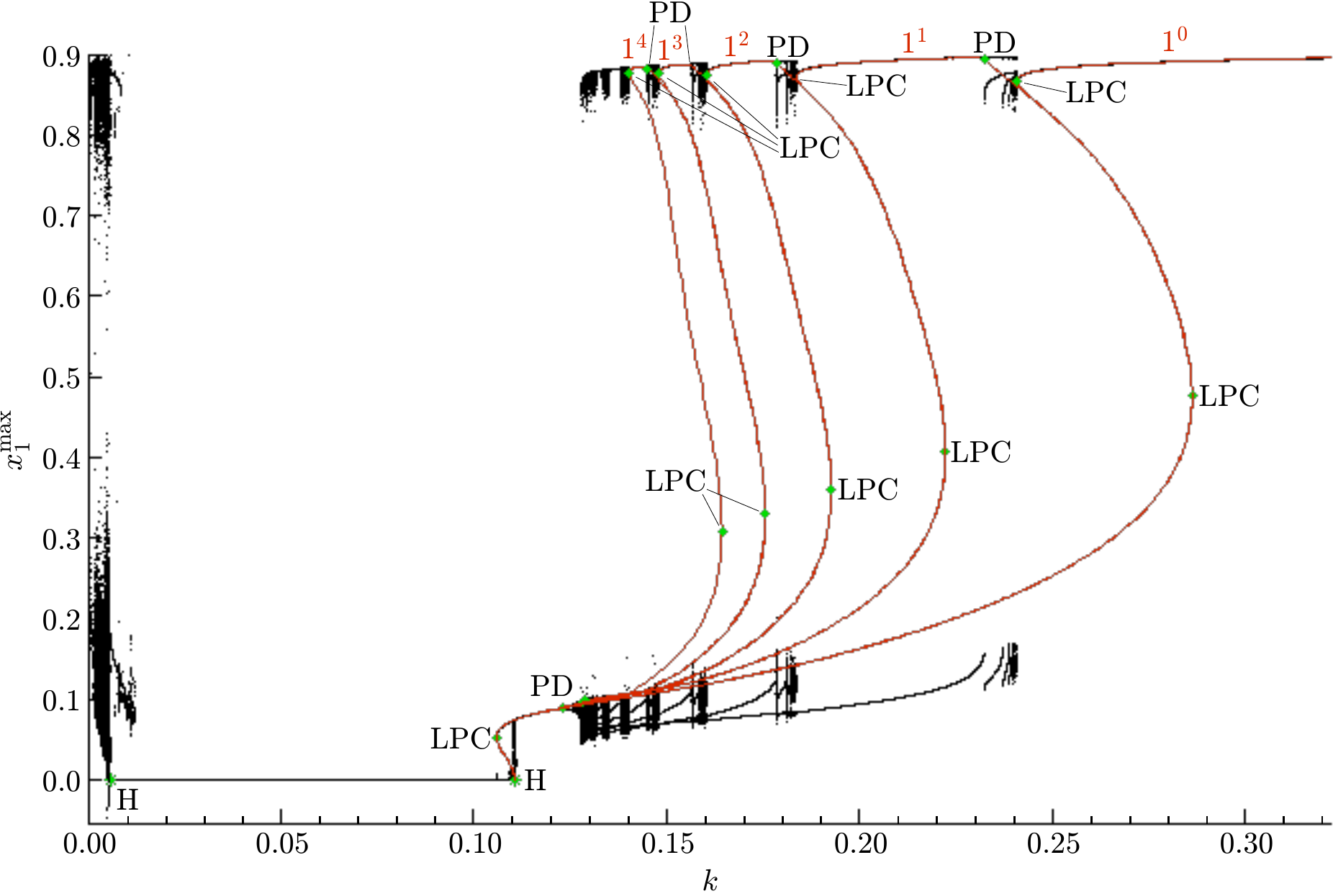}
	\caption{(Color online) Continuation of certain periodic orbits (primary mixed-mode oscillations) and the origin superimposed over the original bifurcation diagram.
	H~marks a Hopf bifurcation, PD a period doubling, and LPC corresponds to the limit point (saddle-node) bifurcation of cycles.
	$L^S$ marks a mixed-mode oscillation with $L$ consecutive high-amplitude and $S$~consecutive low-amplitude oscillations
}
	\label{fig:bifurcation2}
\end{figure*}

\subsection{Mixed-mode oscillations}\label{sec:MMOs}
In the following, we will investigate in more detail the periodic solutions observed in the bifurcation diagram and their characteristics.
Fig.~\ref{fig:bifurcation2} shows results obtained with continuation methods for system~A.
Since the oscillations contain small and large amplitude oscillations, we use the concept of mixed-mode oscillations and denote these mixed-mode forms by the standard $L^S$ notation, where $L$~is the number of consecutive high-amplitude oscillations of $x_1$ and $S$~the number of low-amplitude oscillations~\cite{Desroches2012}.
Note that high-amplitude oscillations of $x_1$ always coincide with high-amplitude-oscillations of $\bar{x}$, $x_2$, $y_1$, and $y_2$.

\subsubsection{Primary mixed-mode oscillations}

As discussed in Sec.~\ref{sec:bifurcation}, the subcritical Hopf bifurcation at $k \approx 0.110687$ gives rise to a period-one limit-cycle attractor, which period-doubles at $k \approx 0.12323$.
Focusing initially on these period-one and period-two limit cycles, we followed their evolution using the continuation software MatCont~\cite{Dhooge2003}.
We observe that the period-one limit cycle continues to exist (although unstable) until $k=0.24158$, at which it stabilizes again through a saddle-node bifurcation of a limit cycle, although with a much higher amplitude.
This period-one limit cycle, which we denote as $1^0$, is stable for higher values of~$k$.
The period-two limit cycle also destabilizes via a period doubling at $k=0.125431$.
Similar to the $1^0$ case, continuation suggests that this attractor continues to exist and stabilizes again at $k\approx 0.183836$ via a saddle-node bifurcation of a limit cycle.
Interestingly, this stabilization yields an attractor with a $1^1$ (one large, one small) mixed-mode oscillation (MMO).
The amplitude of the period-two attractor appears to have gone through a deformation, which causes the difference in amplitude between the two oscillations.
This $1^1$ mixed-mode form is stable for $0.183836 \lkl 0.232397$ before destabilizing through a period-doubling cascade.
The origin of these two periodic solutions was the limit cycle emerging from the subcritical Hopf bifurcation.
It is important to note that the continuation of the period-two solution gives us an isola (a closed curve) in the bifurcation diagram.
A number of other isolas similar to the period-two continuation also exist.

We now focus on the MMO forms of $1^2$, $1^3$, and $1^4$.
For $0.160398 \lkl 0.178387$, we observe a $1^2$~MMO.
Similar to the previous cases, this $1^2$~MMO stabilizes through a saddle-node bifurcation of a limit cycle at $k\approx 0.16039$ and destabilizes via period doubling at $k\approx 0.178387$.
Continuation suggests that this $1^2$~MMO is actually an evolved period-three solution, which is stable for $0.128177 \lkl 0.128227$ and destabilizes via period doubling.
The $1^3$~and $1^4$ mixed-mode oscillations seem to have a similar origin, the former being a stable distortion of a period-four solution existing for $0.12868 \lkl 0.128685$ and the latter of a period-five solution existing for $0.128873 \lkl 0.128874$, respectively.
This shows that the origin of the higher-order MMOs is quite distinct from the initial $1^1$~and $1^0$~MMOs as the parent periodic solutions of these exist in the post-crisis regime in contrast to the latter.
It is also important to note that the parent periodic solutions of these higher-order MMO forms themselves exhibit no high-amplitude oscillations, but evolve and undergo a deformation to yield the respective MMO forms, similar to the $1^0$ and $1^1$ MMOs.

Moreover, we observe that the coupling regimes for which primary MMOs exist scale as $\Delta k_P \propto P^\rho$~\cite{Kaneko1983}, where $P = L+S$ is the period of the MMO. The scaling exponent is obtained as $\rho \approx -2.8$.

\subsubsection{Secondary mixed-mode oscillations}
Each chaotic window separating the windows with primary MMOs mentioned above contains many small windows with secondary MMOs.
For instance, between the MMOs $1^1$ and $1^0$, sequences of $2^1, 3^1, 4^1, \ldots$ are observed embedded in chaotic parameter regions.
Similarly, between $1^2$ and $1^1$, $1^3$ and $1^2$, $1^4$ and $1^3$, MMO sequences of $2^3, 3^4, 4^5, \ldots$, $2^5, 3^7, 4^9, \ldots$, and $2^7, 3^{10}, 4^{13}, \ldots$ are observed, respectively, separated by chaos.

Let $\klg{L^S_i \middle | i \in \mathbb{N}^+}$ denote the secondary MMO sequence that exists between the primary forms $L^S$ and $L^{S+1}$.
A secondary MMO $L^S_i$ remains stable for a certain coupling range before it destabilizes via a period-doubling cascade leading to a regime of chaos, from which the following stable MMO $L^S_{i+1}$ emerges.
This continues until the secondary MMO states and corresponding chaotic attractors accumulate in a very small coupling range when the saddle-node bifurcation of the primary MMO $L^{S+1}$ occurs.

Importantly, the chaotic regimes observed before the transitions from primary MMOs to secondary MMOs appear to be of mixed-mode-chaos type~\cite{Larter1991, Petrov1992}:
A number of low-amplitude oscillations are interrupted by high-amplitude events aperiodically.
The low-amplitude oscillations exhibited in mixed-mode-chaos (and thus the event rate~$R$) appears to be dictated by the number of low-amplitude oscillations~$S$ of the parent MMOs (see also Fig.~\ref{fig:bifurcation1}).
On increasing the coupling strength, the primary MMO states $1^S$ show a gradual decrease in $S$.
Since the number of low-amplitude oscillations decreases, the corresponding mixed-mode chaos also exhibits extreme-event-like trajectories with increasing regularity before vanishing completely with the stabilization of the $1^0$~MMO form.

Similar to the primary MMOs, we observe that the coupling regimes for which the secondary MMOs exist scale with their period $P$ as $\Delta k_P \propto P^\rho$ with $\rho$ taking an almost identical value as for the primary MMOs.

\subsection{Channel size} \label{sec:channelproperties}
As discussed in Sec.~\ref{sec:bifurcation}, the mechanism which leads to extreme events in system~A is the opening of a channel-like structure in state space, denoted by $\mathcal{C}$ in the following.
In this section, we study the possible variations of the size of $\mathcal{C}$ with changes in the coupling strength.
To get an idea about the size of $\mathcal{C}$, let us consider system~A from the perspective of leaking chaotic systems~\cite{Altmann2013}.
In this description, we consider $\mathcal{C}$ as the source of leak and each extreme event as a leaked trajectory.

For leaking chaotic systems, the probability $p(t)$ that a trajectory survives a leak up to time $t$ can be expressed as
	\begin{align}
		p(t)=p(0) \exp{(-R' t)},
		\label{eq:leakprobability}
	\end{align}
where $R'$ is the escape rate and $p(0)$ is the initial survival probability.
The mean survival time $\bar{\tau}'$ of a trajectory can be expressed as $\bar{\tau}' \defi R'^{-1}$.
Kac's lemma \cite{Kac1959} guarantees that the rate~$R'$ of extreme events is equal to the \textit{measure} $\mu(\mathcal{C})$ of the leak for sufficiently small leaks \cite{Altmann2008, Altmann2009}.

We discussed in Ref.~\cite{Ansmann2013} that the inter-event intervals are mostly exponentially distributed for $k=0.128$; slightly above the crisis point.
Calculations suggest that this observation also holds in all other post-crisis chaotic regimes.
This behavior corresponds to the one described in Eq.~(\ref{eq:leakprobability}) with the event rate~$R$ corresponding to the escape rate~$R'$ in leaking chaotic systems.
This similarity and the aforementioned arguments allow us to quantify $\mathcal{C}$ directly from the inter-event rate $R$.
The bottom row of Fig.~\ref{fig:bifurcation1} shows the variation in $R$ for system~A depending on the coupling strength~$k$.
We observe a general tendency of $R$ to increase with increasing~$k$.
However, we can distinguish between regimes in which $R$ is almost constant and those in which $R$ fluctuates.
The former regimes correspond to the periodic primary MMO dynamics, while the fluctuations of~$R$ in the latter regimes are a result of secondary MMO dynamics and the corresponding chaos appearing together in small coupling windows.
Since exponentially distributed extreme events occur for all post-crisis chaotic regimes, the corresponding $R$~values quantify the size of $\mathcal{C}$ with a variation in~$k$.
The observed increase of~$R$ implies an increase of $\mu(\mathcal{C})$ and hence an increase in the frequency of extreme events, and vice versa.

\begin{figure}
	\includegraphics{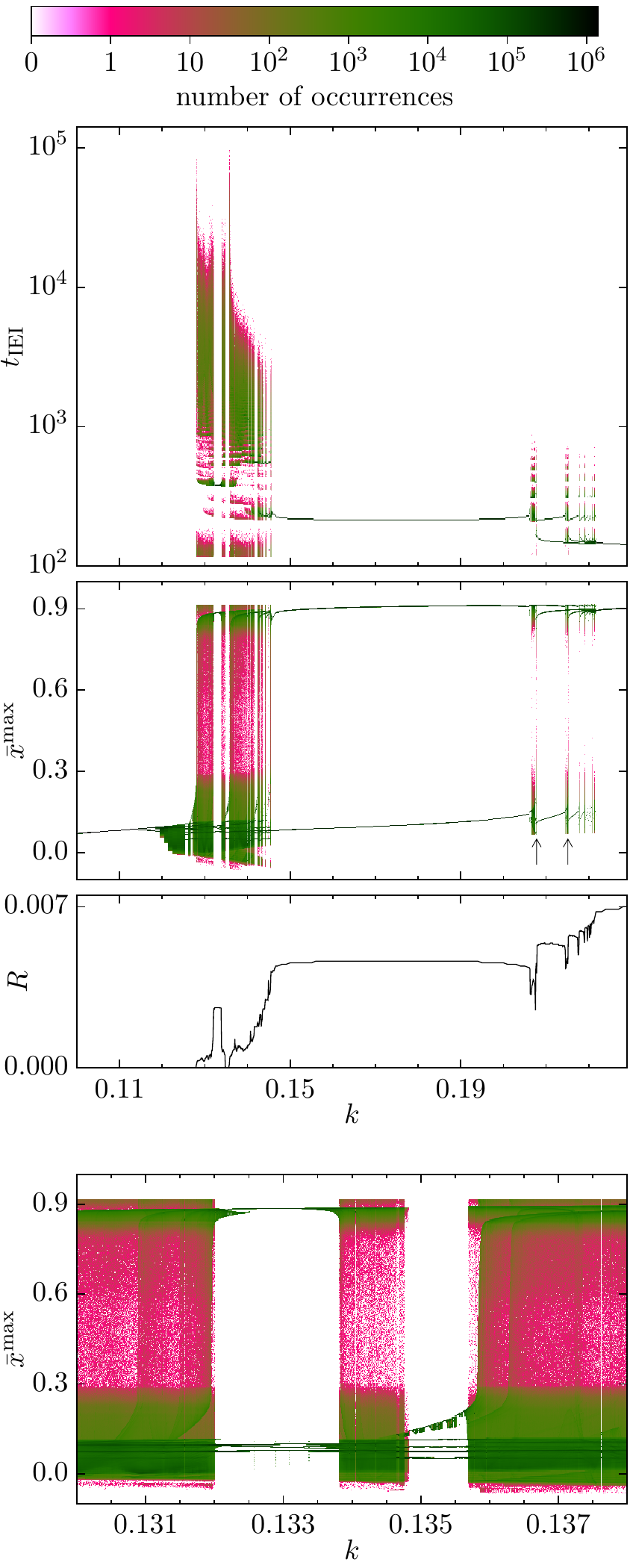}
	\caption{(Color online)
	Same as Fig.~\ref{fig:bifurcation1} but with $10^8$ time units of observation of system~B, with the local maxima of $\bar{x}$ being used in the second row, and with a zoom-in of the second row in the bottom row.
	The two arrows in the second row mark $k=0.2078$ and $k=0.2151$ (see Fig.~\ref{fig:tangent}).}
	\label{fig:bifurcation4}
\end{figure}

\begin{figure}
	\includegraphics{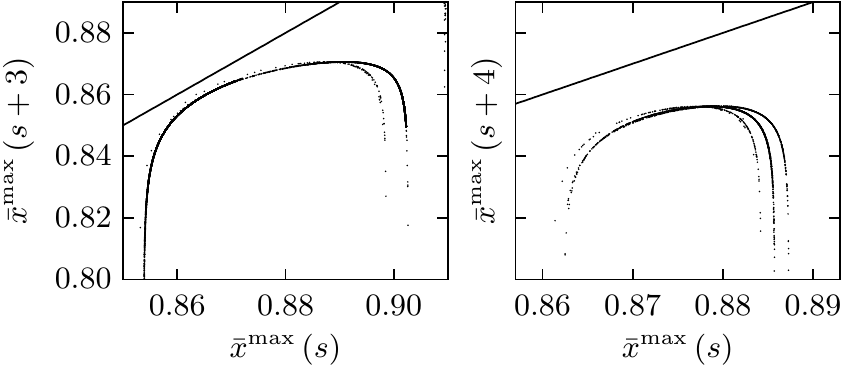}
	\caption{(Left) Third maximum map for $k=0.2078$ in the vicinity of the largest amplitude peak, showing an approaching tangent bifurcation which stabilizes the $2^1$~MMO (see also Fig.~\ref{fig:bifurcation4}).
	(Right) Fourth maximum map showing the approaching tangent bifurcation before the stabilization of the primary $3^1$~MMO at $k=0.2151$ (left).
	The black solid lines show the identity function for reference.}
	\label{fig:tangent}
\end{figure}

\section{System~B}\label{sec:ob.manycoupled}

In this section, we analyze the completely coupled network of 101 FitzHugh--Nagumo units (system~B) to find out whether there is a similar dependence of extreme events on the coupling strength as for system~A.
Particularly we compare the overall dynamics as well as the behavior of the inter-event intervals.
The top two parts of Fig.~\ref{fig:bifurcation4} are bifurcation diagrams for system~B, showing all observed local maxima of $\bar{x}$ and inter-event intervals, depending on the coupling strength $k$.
We observe a sequence of transitions between dynamical states that is comparable to that for system~A and happens on comparable coupling scales:

For $0.08605 \lkl0.115$ we observe period-one oscillations, which originate from a Hopf bifurcation of the origin.
These undergo a period-doubling cascade leading to chaos at $k \approx 0.119$.
This chaotic behavior is interrupted by periodic windows, e.g., around $k \approx 0.127$, and finally undergoes an interior crisis at $k \approx 0.128$ and extreme events appear in the system (cf. Fig.~\ref{fig:timeseries}, third and fourth row).
For $0.128 \lkl 0.145$, we observe a chaotic regime with extreme events, which is interspersed by various MMO forms (see Fig.~\ref{fig:bifurcation4}, bottom). The chaotic region terminates at $k \approx 0.145$ via a period-halving cascade giving rise to a $1^1$~MMO form. This loses stability at $k \approx 0.206$ via a period-doubling cascade leading to chaos again.
There also exists a window around $k \approx 0.135$, in which we observe no high-amplitude oscillations (see Fig.~\ref{fig:bifurcation4}, bottom).
We anticipate that more of such windows exist, and expect that their dynamics evolves to MMOs observed for larger coupling values, which is similar to the case of higher order primary MMOs of system~A.

Although a continuation calculation of MMOs for system~B is difficult, we can still deduce some bifurcation features in the system alternatively.
For the saddle-node bifurcation of specific MMOs, the $m$\textsuperscript{th} maximum map $(m=3,4)$ in the vicinity of the largest amplitude peak in Fig.~\ref{fig:tangent} shows the approaching tangency with the $m$\textsuperscript{th} period solution of this map.
Chaos obtained from period doubling of the $1^1$~MMO terminates at one such tangency (see the left part of Fig.~\ref{fig:tangent}), which leads to a stable $2^1$~MMO.
For larger values of~$k$, we observe similar transitions between chaotic and MMO windows (see also the right part of Fig.~\ref{fig:tangent} for instance), which become smaller and smaller and finally accumulate and terminate at a saddle-node bifurcation of a limit cycle at $k\approx 0.221$ leading to a $1^0$~oscillation.

For $k=0.128$, inter-event intervals are mostly exponentially distributed close to the interior crisis~\cite{Ansmann2013} (except for a certain clustering of extreme events, see also Fig.~\ref{fig:bifurcation4}, top).
This kind of distribution can be observed for all other investigated chaotic windows between $k \approx 0.128$ and $k \approx 0.145$ (see, e.g., the approximately constant number of occurrences in each of the logarithmically sized bins in the top of Fig.~\ref{fig:bifurcation4} for $10^3<t_\text{IEI}<10^4$ and $0.128<k<0.132$).
Together with considerations similar to those for system~A (see Sec.~\ref{sec:channelproperties}), we can thus consider system~B in analogy to leaking chaotic systems and regard the event rate~$R$ as the measure of the leak.
As for system~A, we observe the event rate~$R$ (see Fig.~\ref{fig:bifurcation4}, third row) to tendentially increase with increasing~$k$, however, increases mostly happen in the chaotic regimes, while $R$ slightly decreases in the MMO regimes.

\begin{figure}
	\includegraphics{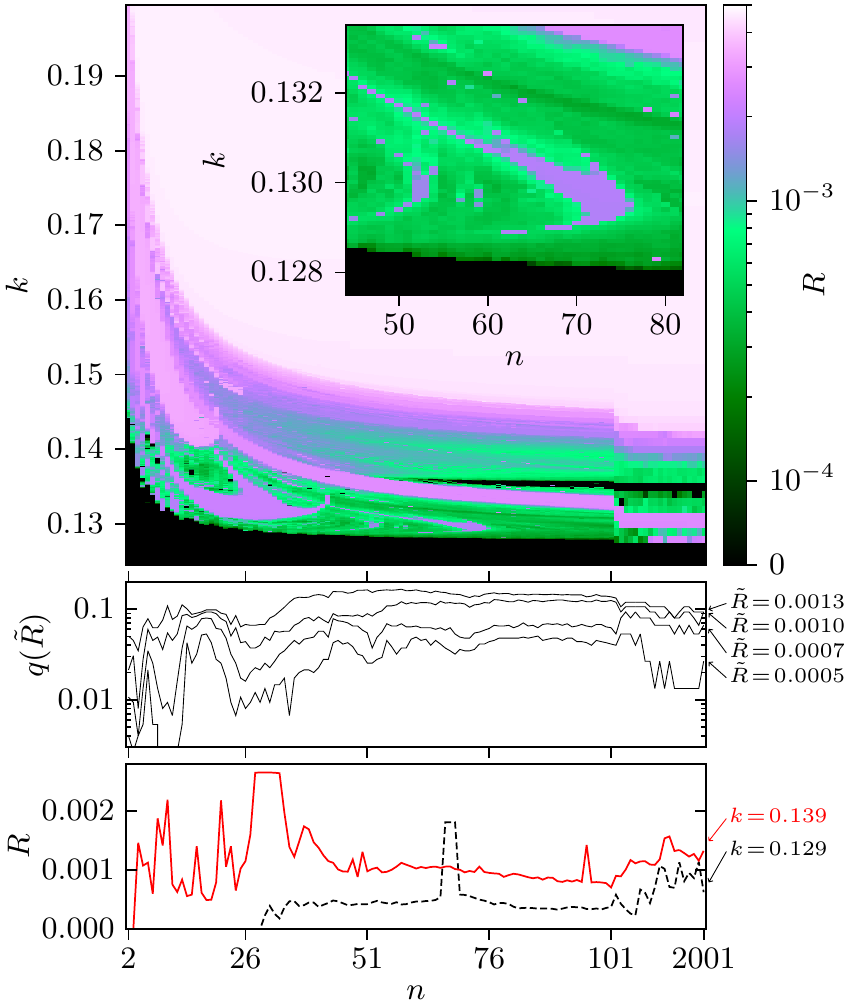}
	\caption{(Color online)
	(Top) Event rate $R$ (color-coded) depending on the number of oscillators~$n$ and on the normalized coupling strength~$k$ for generalizations of system~B, using the same parameters.
	(Middle) Estimated fraction $q(\tilde{R})$ of $k$~values with $0<R<\tilde{R}$ for several thresholds~$\tilde{R}$, i.e., $q(\tilde{R}) \defi \int_{0.125}^{0.200} \Theta(R(k))\Theta(\tilde{R}-R(k)) \diff k$, where $\Theta$ denotes the Heaviside function and $\Theta(0) \defi0$.
	For $n>101$, the scaling of the abscissa is centupled, and the resolution of the ordinate and thus the accuracy of $q(\tilde{R})$ is reduced.
	(Bottom) $R$ in dependence of $n$ for selected $k$: black, dashed line: $k=0.129$; red, solid line: $k=0.139$.
}
	\label{fig:n_dep}
\end{figure}

\section{Dependence on system size} \label{sec:dependence_systemsize}

Comparing the results for systems A and~B, we see that the coupling regimes in which rare events can be observed are much larger for system~B, i.e., for the larger system (cf., the top rows of Figs. \ref{fig:bifurcation1} and~\ref{fig:bifurcation4}).
To better understand this dependence on system size, we investigate the generalizations of system~B to network sizes $n \in \klg{2, 3, \ldots, 101} \cup \klg{101, 201, \ldots, 2001}$, whose event rates~$R$ depending on the coupling strength~$k$ are shown in Fig.~\ref{fig:n_dep}.

Overall, for a fixed $k$, we observe $R$ to undergo strong changes for small~$n$ and to behave asymptotically for large~$n$.
In particular, we observe the fraction of $k$~values with rare events~($q$) to vary strongly, but tendentially increase until $n\approx 40$ and exhibit a behavior that is between slightly decreasing and constant for larger~$n$.
We thus expect that rare events continue to be observable in a comparatively large coupling regime for at least several orders of magnitude of~$n$.
In the bottom of Fig.~\ref{fig:n_dep}, we show the event rate~$R$ for selected, fixed~$k$.
Apart from some high and almost constant plateaus of the event rate~$R$ due to periodic dynamics, we can regard it as a measure of the leak.
We observe that, after some initial fluctuations for small $n$, the leak measure remains largely constant when the system size~$n$ is increased.

If we consider the system size~$n$ as a control parameter, we observe structures that resemble shrimps \cite{Gallas1993, Stoop2012}, e.g., around $k = 0.133$ for about $15<n<35$ and around $k = 0.130$ for about $65<n<75$ (see Fig.~\ref{fig:n_dep}).

\section{Conclusions}\label{sec:conclusions}
In this paper, we investigated various dynamical regimes and transitions between them in one system consisting of~2 and one network consisting of 101 diffusively coupled inhomogeneous FitzHugh--Nagumo units.
These systems are capable of generating extreme events for certain parameter values and we observe them to emerge after an interior crisis, which is similar to the observations made for rogue events in a deterministic optical system in Ref.~\cite{ZamoraMunt2013}.
Performing a bifurcation analysis, we observed various post-crisis mixed-mode-oscillatory (MMO) regimes, whose extension in parameter space we observed to follow a power law with respect to the period for the two-unit system.
For both systems, these MMOs are mediated by chaotic bands, in which we observe extreme events, and stabilize via a saddle-node bifurcation of a limit cycle.

The emergence of extreme events is related to the opening of a channel-like structure through which the trajectory can escape to form an event.
This opening of the channel can be studied by computing the stable and unstable manifolds of the fixed point associated with the mechanism of the emergence.
We found that the size of the channel-like structure increases with increasing distance from the crisis point.
However, such a computation is impossible for higher-dimensional systems such as the network.
Therefore we have employed the concept of leaking chaotic systems and estimated the size of the channel in a high-dimensional state space using the relationship between the measure of the channel-like structure, i.e., the leak, and the mean inter-event rate as given by Kac's lemma.
This allows us to study geometric properties of manifolds in a high-dimensional state space.
One would expect that these properties change with the number of dimensions involved, i.e., with the number of nodes of the network.
It turned out that the size of the channel is almost independent of the number of nodes, at least for the number of nodes investigated here.
We conjecture that this is due to the fact that we consider only networks with global coupling, so that adding new nodes does not extend the complexity of the topological structure.
For other coupling topologies, we expect a more complicated dependence of the size of the channel on the network size.

Our observations suggest that, despite their different complexities and dynamical properties, the two systems have a similar bifurcation structure.
However, we observe that the regimes of chaotic behavior containing extreme events are much wider for the larger system.
Analyzing systems whose sizes ranged several orders of magnitude, we found that the width of the extreme-event regimes and hence the robustness of the phenomenon to first tendentially increase and then remain roughly stable with system size.
We note that the bifurcations in the larger system are very well represented by the map constructed by the local maxima of the mean value of the first variable ($\bar{x}^\text{max}$) for sufficiently high coupling.
We expect that to be true for any system size.

To summarize, we investigated in detail how extreme events are generated in systems of coupled FitzHugh--Nagumo units of various sizes.
Since such units are widely used for modeling natural excitable systems, our findings can contribute to increasing our understanding as to how extreme events emerge from their dynamics.

\section*{Acknowledgments}
The authors would like to thank R.E.~Amritkar, S.~Bialonski, C.~Kuehn, A.~Rothkegel, and T.~T\'el for interesting discussions.
We are grateful to S.~Bialonski and A.~Rothkegel for critical comments on earlier versions of the manuscript.
This work was supported by the Volkswagen Foundation (Grant~Nos. 85388 and 85392).
U.F.~would like to thank R.~Roy and his group for their hospitality and the Burgers Program for Fluid Dynamics of the University of Maryland for financial support.

\appendix

\begin{figure*}
	\includegraphics{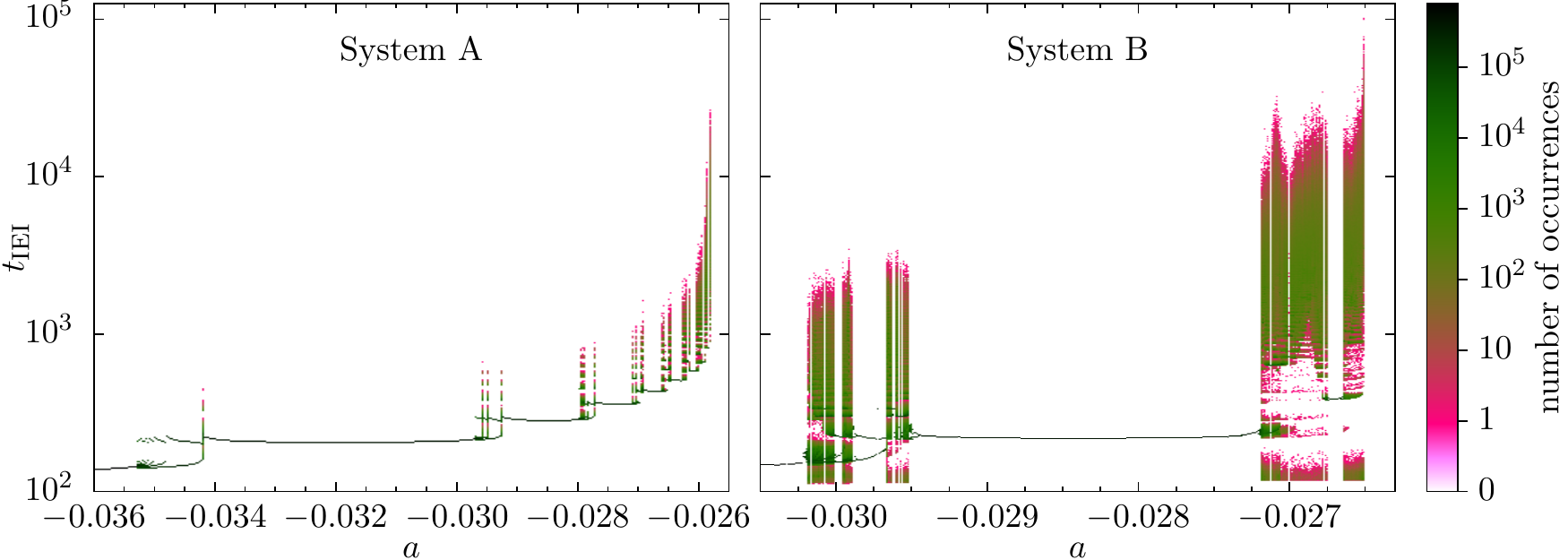}
	\caption{(Color online)
	(Left) Same as the top row of Fig.~\ref{fig:bifurcation1}, but with $k$ fixed to $0.128$ and $a$ varied.
	(Right) Same as the top row of Fig.~\ref{fig:bifurcation4}, but with $k$ fixed to $0.128$ and $a$ varied.}
	\label{fig:bifurcation_a}
\end{figure*}

\section{Dependence on internal parameters} \label{sec:internal}

In Fig.~\ref{fig:bifurcation_a} we show bifurcation diagrams for systems A and~B in dependence of the internal parameter~$a$ (cf.~Eq.~\ref{eq:systems}).
Apart from the flipped sign, we observe similar transitions between dynamical regimes as for varying the coupling strength~$k$.
In particular, we still have the Hopf bifurcation of the origin giving rise to a period-one attractor, chaos emerging from a period-doubling cascade of this period-one limit cycle, the chaotic attractor undergoing an interior crisis beyond which we obtain extreme events, and the intermediate MMOs separating different chaotic bands in the post-crisis regime.
The biggest difference to our results for varying~$k$ lies in the transition of system~B from the $1^1$~MMO to the $1^0$~oscillation, for which the ratio between chaotic and periodic windows as well as the maximal inter-event intervals are larger for varying~$a$ emphasizing the parameter regimes in which events occur.

We observe similar results for varying the remaining parameters of the units or systems, respectively, namely for~$c$, for the mean of the~$b_i$, and in case of system~B for the spread of the~$b_i$.

\section{Details of transitions for small coupling strengths in system~A} \label{sec:detailsoftransitions}
\begin{figure}
	\includegraphics{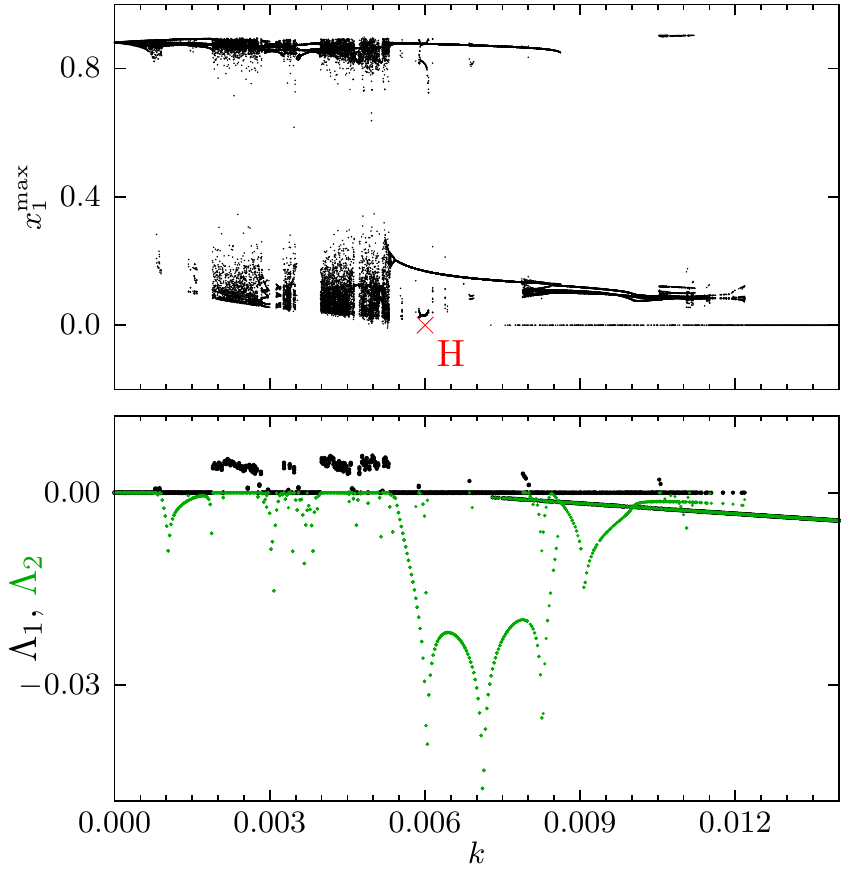}
	\caption{(Top) Zoom into the bifurcation diagram shown in Fig.~\ref{fig:bifurcation2}.
	H~marks a Hopf bifurcation.
	(Bottom) Dependence of the two largest Lyapunov exponents $\Lambda_1$ (black) and~$\Lambda_2$ (green) on the coupling strength~$k$.}
	\label{fig:zoomed}
\end{figure}
Figure \ref{fig:zoomed} shows the bifurcation diagram of the system for $k < 0.014$.
We observe that with an increase of the coupling strength~$k$, the system follows a quasiperiodic route to a chaotic state at $k \approx 0.00075$.
This chaotic state is interrupted by periodic windows and persists for $0.00075 \lkl 0.000916$.
For larger values of~$k$, the system exhibits multistability between a period-five limit cycle and chaos.
Nevertheless, the Lyapunov exponents suggest that the system is quite close to quasiperiodicity for most of this regime.
Additionally, the basin of attraction for chaos expands with increasing $k$, and consequently the basin for the periodic behavior shrinks and disappears at $k \approx 0.00188637$, beyond which only the chaotic attractor exists.
For larger values of~$k$, periodic windows appear in the system via period-halving cascades and disappear again via period-doubling bifurcations leading to chaos again.
At $k \approx 0.00541$, the chaotic regime terminates via period halving, giving rise to a period-two limit cycle which exists thereafter.
The system exhibits bistability between periodic--periodic and periodic--chaotic attractors in small parameter ranges with an increase in $k$.
In this parameter range, chaos seems to appear abruptly but this might very well be a projection effect of analyzing only one of four dynamical variables.
This chaotic attractor period-halves, leading to periodic behavior and then these periodic orbits period-double and give rise to chaos again (e.g., for $0.00587 \lkl 0.00607$).

Additionally, at $k \approx 0.006$, the equilibrium at the origin stabilizes via a reverse Hopf bifurcation leading to multistability.
Although the Hopf bifurcation happens at $k \approx 0.006$, we observe the corresponding fixed-point dynamics only for $k \approx 0.00726$.
This can probably be related to the small basin of attraction of the origin for values of~$k$ close to the Hopf bifurcation and to the increasing basin size with increasing~$k$.

For $k \gtrapprox 0.00726$, the system exhibits multistability between fixed-point, periodic (period~2), chaotic attractors ($0.00789 \lkl 0.00803$) and fixed-point, periodic (period~2), and periodic attractors ($0.00803 \lkl 0.00862$).
The period-two behavior is the one which came into existence at $k \approx 0.00541$ and the other periodic behavior emerges from the chaos for $0.00789 \lkl 0.00803$ via period halving which finally settles on a period-three behavior.
The period-two attractor vanishes at $k \approx 0.00862$ via a saddle-node bifurcation of a limit cycle, leaving the system in a bistable state with coexisting period-three and fixed-point attractors.
Multistability reappears with the appearance of chaos in certain small parameter ranges.
These chaotic regimes vanish via period halving yielding periodic states.
The system shows several such appearances of chaos and periodic behaviors leading to complicated multistability scenarios.
For $k \gtrapprox 0.0121$, the period-three attractor period-doubles and leads to chaos, which then terminates abruptly at $k \approx 0.01218$.

\end{document}